# Spectroscopic evidence for bulk-band inversion and three-dimensional massive Dirac fermions in ZrTe$_5$

Zhi-Guo Chen[a,1], R. Y. Chen[b], R. D. Zhong[c], John Schneeloch[c], C. Zhang[c], Y. Huang[d], Fanming Qu[a,e], Rui Yu[f], Q. Li[c], G. D. Gu[c] and N. L. Wang[b,g,1]

[a]Beijing National Laboratory for Condensed Matter Physics and Institute of Physics, Chinese Academy of Sciences, Beijing 100190, China

[b]International Center for Quantum Materials, School of Physics, Peking University, Beijing 10087, China

[c]Condensed Matter Physics and Materials Science Department, Brookhaven National Lab, Upton, New York 11973, USA

[d]Center for Functional Nanomaterials, Brookhaven National Lab, Upton, New York 11973, USA

[e]QuTech, Delf University of Technology, Delf 2600 GA, The Netherlands

[f]School of Physics and Technology, Wuhan University, Wuhan 430072, China

[g]Collaborative Innovation Center of Quantum Matter, Beijing, China

[1]To whom correspondence may be addressed. E-mail: zgchen@iphy.ac.cn or nlwang@pku.edu.cn

**Three-dimensional topological insulators (3D TIs) represent novel states of quantum matters in which surface states are protected by time-reversal symmetry and an inversion occurs between bulk conduction- and valence-bands. However, the bulk-band inversion which is intimately tied to the topologically nontrivial nature of 3D TIs has rarely been investigated by experiments. Besides, 3D massive Dirac fermions with nearly near band dispersions were seldom observed in TIs. Recently, a van der Waals crystal, $ZrTe_5$, was theoretically predicted to be a TI. Here, we report an infrared transmission study of a high-mobility (~ 33,000 $cm^2/(V·s)$) multilayer $ZrTe_5$ flake at magnetic fields ($B$) up to 35 T. Our observation of a linear relationship between the zero-magnetic-field optical absorption and the photon energy, a bandgap of ~ 10 meV and a $\sqrt{B}$-dependence of the Landau level (LL) transition energies at low magnetic fields demonstrates 3D massive Dirac fermions with nearly linear band dispersions in this system. More importantly, the reemergence of the intra-LL transitions at magnetic fields higher than 17 T reveals the energy cross between the two zeroth LLs, which reflects the inversion between the bulk conduction- and valence-bands. Our results not only provide spectroscopic evidence for the TI state in $ZrTe_5$ but also open up a new avenue for fundamental studies of Dirac fermions in van der Waals materials.**

## Significance

Experimental verifications of the theoretically predicted topological insulators (TIs) are essential steps towards the applications of the topological quantum phenomena. In the past, theoretically predicted TIs were mostly verified by the measurements of the topological surface states. However, as another key feature of the nontrivial topology in TIs, an inversion between the bulk-bands has rarely been observed by experiments. Here, by studying the optical transitions between the bulk LLs of $ZrTe_5$, we not only offer spectroscopic evidence for the bulk-band inversion—the crossing of the two zeroth LLs in a magnetic field but also quantitatively demonstrate three-dimensional massive Dirac fermions with nearly linear band dispersions in $ZrTe_5$. Our investigation provides a new paradigm for identifying TI states in candidate materials.

## Introduction

Topologically nontrivial quantum matters, such as topological insulators (1-8), Dirac semimetals (9-19) and Weyl semimetals (20-27), have sparked enormous interest owing both to their exotic electronic properties and potential applications in spintronic devices and quantum computing. Therein, intrinsic topological insulators have insulating bulk states with odd $Z_2$ topological invariants and metallic surface or edge states protected by time-reversal symmetry (4-6, 28). Most of the experimental evidence to date for TIs is provided by the measurements of the spin texture of the metallic surface states. As a hallmark of the nontrivial $Z_2$ topology of TIs (4-6, 28), an inversion between the characteristics of the bulk conduction- and valence-bands occurring at an odd number of time-reversal invariant momenta has seldom been probed by experiments. An effective approach for identifying the bulk-band inversion in TIs is to follow the evolution of two zeroth Landau levels (LLs) which arise from the bulk conduction- and valence-bands, respectively. As shown in Fig. 1*A*, for TIs, due to the bulk-band inversion and Zeeman effects, the two zeroth bulk Landau levels are expected to intersect in a critical magnetic field and then separate (3, 29), while for trivial insulators, the energy difference between their two zeroth Landau levels would become larger with increasing magnetic field. Therefore, an intersection between the two zeroth bulk LLs is a significant signature of the bulk-band inversion in TIs. However, a spectroscopic study of the intersection between the two zeroth bulk LLs in 3D TIs is still missing. In addition, many typical 3D TIs, such as $Bi_2Se_3$, show massive bulk Dirac fermions with parabolic band dispersions, which are effectively described by massive Dirac models (6, 28, 29). By contrast, 3D massive Dirac fermions with nearly linear bulk band dispersions (7), which are interesting topics following 2D massive Dirac fermions in gapped graphene (30, 31), were rarely observed in 3D TIs.

A transition-metal pentatelluride, $ZrTe_5$, embodies both 1D chain and 2D layer features (32), shown in Fig. 1*B*. One Zr atom together with three Te (1) atoms forms a quasi-1D prismatic chain $ZrTe_3$ along the *a*-axis (*x*-axis). These prismatic $ZrTe_3$ chains are connected through zigzag chains of Te (2) atoms along the *c*-axis (*y*-axis) and then construct quasi-2D $ZrTe_5$ layers. The bonding between $ZrTe_5$ layers is van der Waals type (33, 34). Thus, as displayed in Fig. 1*C*, bulk $ZrTe_5$ crystals can be easily cleaved down to a few layers. Recently, the *ab initio* calculations indicate that monolayer $ZrTe_5$ sheets are great contenders for quantum spin Hall insulators—2D TI and that 3D $ZrTe_5$ crystals are quite close to the phase boundary between strong and weak TIs (33). Scanning tunneling microscopy measurements have shown that edge states exist at the step edges of the $ZrTe_5$ surfaces (35, 36). Nonetheless, further investigations are needed to check whether the observed edge states in $ZrTe_5$ are topologically nontrivial or not. Studying the bulk-band inversion or the intersection between the two zeroth bulk LLs can provide a crucial clue to clarifying the nature of the edge states in $ZrTe_5$. Except the edge states within the energy gap of the bulk bands around the Brillouin zone center (i.e. Γ point) of $ZrTe_5$ (36, 37), 3D massless Dirac fermions with the linearly dispersing conduction- and valence-band degenerate at the Γ point were suggested to exist in this material by previous angle-resolved photon emission spectroscopy, transport and optical experiments (38-41). Considering that (1) our $ZrTe_5$ thick crystals were experimentally shown to be Dirac semimetals hosting 3D massless Dirac fermions, (2) $ZrTe_5$ monolayers were theoretically predicted to be quantum spin Hall insulators and (3) the bulk state of $ZrTe_5$ is very sensitive to its interlayer distance which might be discrepant in different samples (33, 40), it is significant to quantitatively verify whether 3D massive Dirac fermions with a bandgap and nearly linear bulk-band dispersions can be realized in dramatically thinned flakes of our $ZrTe_5$ crystals.

Infrared spectroscopy is a bulk-sensitive experimental technique for studying low-energy excitations of a material. Here, in order to investigate the bulk-band inversion and the nature of the bulk fermions in $ZrTe_5$, we measured the infrared transmission spectra $T(\omega, B)$ of its multilayer flake with thickness $d \sim$ 180 nm at magnetic fields applied along the wave vector of the incident light (see Materials, Methods and Supplementary Information Section 1). A series of intra- and inter-LL transitions are present in the relative transmission spectra $T(B)/T(B_0 = 0 \text{ T})$ of the $ZrTe_5$ flake. The linear $\sqrt{B}$-dependence of the LL transition energies at $B \leq 4$ T and the non-zero intercept of the LL transitions at $B = 0$ T, combined with the linear relationship between the zero-magnetic-field optical absorption and the photon energy, indicates 3D massive Dirac fermions with nearly linear band dispersions in the $ZrTe_5$ flake. Moreover, a 3D massive Dirac model with a bandgap of $\sim 10$ meV can quantitatively explain the magnetic-field dependence of the measured LL transition energies very well. At high magnetic fields, we observed fourfold splittings of the LL transitions. Additionally, our analysis of the split LL transitions shows that the intra-LL transitions, which are associated to the two zeroth LLs and disappear at $B \approx 2.5$ T, reemerge at $B > 17$ T. Considering that the zeroth LL crossing in a Zeeman field would make the two zeroth bulk LLs intersect with the chemical potential here and then alter the carrier occupation on the zeroth LLs, we attribute the reemergence of the intra-LL transitions in the $ZrTe_5$ flake to the energy crossing of its two zeroth bulk LLs which originates from the bulk-band inversion. These results strongly support the theoretically predicted 3D TI states in 3D $ZrTe_5$ crystals.

## Results

**3D massive Dirac fermions.** At zero magnetic field, the measured absolute transmission $T(\omega)$ corresponds to the absorption coefficient: $A(\omega) = -[\ln T(\omega)]/d$, where $d$ is the thickness of the sample and $\omega$ is the photon energy (see the Methods of Ref. 42). In solids, the absorption coefficient is determined by the joint density of state $D(\omega)$: $A(\omega) \propto D(\omega)/\omega$. 3D electron systems with linear band dispersions along three momentum directions have the $D(\omega)$ proportional to $\omega^2$, while for 2D linear dispersions, $D(\omega) \propto \omega$. Thus, in stark contrast to the $\omega$-independent absorption of 2D Dirac materials like graphene (43), the linear $\omega$-dependence of $A(\omega)$ in Fig. 1*D* indicates 3D linear band dispersions in $ZrTe_5$. Moreover, at low energies, the absorption coefficient apparently deviates from the linear relationship with $\omega$, implying the opening of a bandgap at the original Dirac point. The 3D linear band dispersions, together with the possible bandgap, suggest the presence of 3D massive Dirac fermions in the exfoliated ZrTe$_5$ flake.

In order to confirm the 3D massive Dirac fermions in ZrTe$_5$, we further performed infrared transmission experiments at magnetic fields applied perpendicular to the *ac*-plane (*xy*-plane) of the crystal (Faraday geometry). The low-field relative transmission $T(B)/T(B_0 = 0\text{ T})$ spectra in Fig. 2*A* show seven dip features T$_n$ ($1 \leq n \leq 7$) directly corresponding to the absorption peaks of LL transitions. All of these dip features systematically shift to higher energies, as the magnetic field increases. Here, we define the energy positions of the transmission minima in $T(B)/T(B_0)$ as the absorption energies, which is a usual definition in thin film systems, such as Bi$_2$Se$_3$ films and graphene. Then, we plotted the square of the T$_n$ energies ($E^2_{T_n}$) as a function of magnetic field in Fig. 2*B*. The linear *B*-dependence of $E^2_{T_n}$ (i.e. linear $\sqrt{B}$-dependence of $E_{T_n}$) reveals the LL transitions of Dirac fermions (31, 44, 45). In Fig. 2*C*, the non-zero intercept of the linear fit to $E^2_{T_1}$ at zero magnetic-field is an important signature of a non-zero Dirac mass or a bandgap (31). Therefore, the linear relationship between $E^2_{T_n}$ and *B*, together with the non-zero intercept, provides further evidence for 3D massive Dirac fermions in the ZrTe$_5$ flake.

In order to quantitatively check the 3D massive Dirac fermions in the ZrTe$_5$ flake, we employ a 3D massive Dirac Hamiltonian which was derived from the low-energy effective *k · p* Hamiltonian based on the spin-orbital coupling, the point group and time-reversal symmetries in ZrTe$_5$ and includes the spin degree of freedom (40). According to the 3D massive Dirac Hamiltonian expanded to the linear order of momenta, we can obtained the nearly linear band dispersions of ZrTe$_5$ at zero magnetic field: $E(k_{x,y,z}) = \pm\sqrt{\hbar^2(k_x^2 v_x^2 + k_y^2 v_y^2 + k_z^2 v_z^2) + (\Delta/2)^2}$, where $k_{x,y,z}$ are the wave vectors in momentum space, $v_{x,y,z}$ are the Fermi velocities along three momentum directions, Δ is the bandgap and describes the mass of Dirac fermions $m^D_{x,y,z} = \Delta/(2v^2_{x,y,z})$ and $\hbar$ is Planck's constant divided by $2\pi$. In a magnetic field applied perpendicular to the *ac*-plane, the LL spectrum of ZrTe$_5$ without considering Zeeman effects has the form:

$$E_N(k_z) = \pm\delta_{N,0}\sqrt{(\Delta/2)^2 + (v_z\hbar k_z)^2} + \text{sgn}(N)\sqrt{2e\hbar v_F^2 B|N| + (\Delta/2)^2 + (v_z\hbar k_z)^2}, \quad (1)$$

where $v_F$ is the effective Fermi velocity of LLs, the integer *N* is Landau index, $\delta_{N,0}$ is the Kronecker delta function, sgn(*N*) is the sign function and *e* is the elementary charge. The $\sqrt{B}$ dependence is a hallmark of Dirac fermions (31, 44, 45). According to Eq. (1), the magnetic field makes the 3D linear band dispersions evolve into a series of 1D non-equally-spaced Landau levels (or bands) which disperse with the momentum component along the field direction. Specifically, since the 3D massive Dirac Hamiltonian of ZrTe$_5$ involves the spin degree of freedom of this system, two zeroth LLs indexed by $N = +0$ and $N = -0$ locate at the hole and electron band extrema, respectively and have energy dispersions $E_{\pm 0}(k_z) = \pm\sqrt{(\Delta/2)^2 + (v_z \hbar k_z)^2}$, which is different from the case that when the spin degree of freedom was not considered in materials with the hexagonal lattice, only one zeroth non-degenerate Landau level is present in each valley (46). The optical selection rule for ZrTe$_5$ only allows the LL transitions from LL$_N$ to LL$_{N'}$: $\Delta N = |N| - |N'| = \pm 1$ and with the $k_z$-momentum difference $\Delta k_z = 0$ (40). Due to the singularities of the density-of-states (DOS) at $k_z = 0$, magneto-optical response here, which is determined by the joint DOS, should be mainly contributed by the LL transitions at $k_z = 0$ (29, 40). Thus, the energies of the inter-band LL (inter-LL) transitions LL$_{-|N|}$ → LL$_{+|N-1|}$ (or LL$_{-|N-1|}$ → LL$_{+|N|}$) and the intra-band LL (intra-LL) transitions LL$_{+|N-1|}$ → LL$_{+|N|}$ (or LL$_{-|N|}$ → LL$_{-|N-1|}$), $E_N^{Inter}$ and $E_N^{Intra}$ at $k_z = 0$, are given by:

$$E_N^{Inter} = \sqrt{2e\hbar v_F^2 B|N| + (\Delta/2)^2} + \sqrt{2e\hbar v_F^2 B|N-1| + (\Delta/2)^2} \qquad (2)$$

$$E_N^{Intra} = \sqrt{2e\hbar v_F^2 B|N| + (\Delta/2)^2} - \sqrt{2e\hbar v_F^2 B|N-1| + (\Delta/2)^2} \qquad (3)$$

From T$_1$ to T$_7$, the slopes of the linear fits to $E_{T_n}^2$ in Fig. 2B scale as 1 : 5.7 : 9.3 : 13.1 : 16.7 : 20.5 : 24.1, respectively, which is close to the approximate ratio of the theoretical inter-LL transition energies based on Eq. (2), 1 : $(\sqrt{2}+\sqrt{1})^2$ ($\approx$ 5.8) : $(\sqrt{3}+\sqrt{2})^2$ ($\approx$ 9.9) : $(\sqrt{4}+\sqrt{3})^2$ ($\approx$ 13.9) : $(\sqrt{5}+\sqrt{4})^2$ ($\approx$ 17.9) : $(\sqrt{6}+\sqrt{5})^2$ ($\approx$ 21.9) : $(\sqrt{7}+\sqrt{6})^2$ ($\approx$ 25.9). Therefore, the absorption features T$_n$ are assigned as the inter-LL transitions: LL$_{-|N-1|}$ → LL$_{+|N|}$ (or LL$_{-|N|}$ → LL$_{+|N-1|}$) (Fig. 2B) and we have $n = |N|$, where $1 \leq n \leq 7$. Fitting $E_{T_n}^2$ based on Eq. (2) from a least square fit yields a bandgap $\Delta \approx 10 \pm 2$ meV, the effective Fermi velocities $v_F^{T_1} \approx (4.76 \pm 0.04) \times 10^5$ m/s and $v_F^{T_{2 \leq n \leq 7}} \approx (5.04—4.95 \pm 0.04) \times 10^5$ m/s (see Supplementary Section 2).

As another signature of the bandgap or the non-zero Dirac mass, the absorption feature T$_1^*$ is present at energies lower than the lowest-energy inter-LL transition T$_1$ in Fig. 2D. The feature T$_1^*$ is attributed to the intra-LL transition LL$_{+0}$ → LL$_{+1}$ (or LL$_{-1}$ → LL$_{-0}$), illustrated by the grey arrows in Fig. 2E (see Supplementary Section 3). According to Eq. (2) and (3), the energy difference ($E_{T1} - E_{T1*}$) between the transitions T$_1$ and T$_1^*$ in the inset of Fig. 2C directly gives the bandgap value $\Delta = E_{T1} - E_{T1*} \approx 10 \pm 2$ meV, which is consistent with the value obtained by the above fitting. Furthermore, the field dependence of the T$_1^*$ energy in Fig. 2C can be well fitted by Eq. (3) with $\Delta \approx 10 \pm 2$ meV and $v_F^{T_1^*} \approx (4.63 \pm 0.04) \times 10^5$ m/s.

The carrier-charge mobility $\mu$ in the ZrTe$_5$ flake can be calculated using the general formula (47): $\mu = e\hbar/(\Gamma m^*)$, where $\Gamma$ is the transport scattering rate and $m^*$ is the carrier effective mass on the anisotropic Fermi surface (48, 49). Here, the transport scattering rate $\Gamma$ within the $ac$-plane can be roughly estimated from the width of the T$_1$ feature at low fields: $\Gamma \approx 9$ meV at $B = 0.5$ T. Moreover, considering the absence of Pauli blocking of the T$_1$ transition at $B = 0.5$ T, we get the Fermi energy in ZrTe$_5$, $E_F < |E_{LL+1 \text{ (or} -1)}| = E_{T1} \approx 15$ meV (see Supplementary Section 4), which means the Fermi level in our sample is quite close to the band extrema. In this case, the average effective mass $m^*$ of the carriers within the $ac$-plane can be described by (30): $m_{ac}^* \approx \Delta/[2(v_F^{ac})^2] \approx 3.54 \times 10^{-33}$ kg $\approx 0.00389$ $m_0$, where $m_0$ is the free electron mass and the average Fermi velocity within the $ac$-plane $v_F^{ac}$ is approximately equal to the effective Fermi velocity of the LLs, $v_F^{ac} \approx 4.76 \times 10^5$ m/s. Finally, we can estimate the mobility of the carriers within the $ac$-plane of our ZrTe$_5$ sample: $\mu \approx 33{,}000$ cm$^2$/(V·s), which is comparable to those in graphene/h-BN heterostructures (50, 51).

**Buk-band Inversion.** As shown in Fig. 3*A*, applying a higher magnetic field enables us to observe the splitting of the T$_1$ transition, which indicates a non-negligible Zeeman effect in ZrTe$_5$ (40). For TIs, due to the Zeeman field, each LL except the two zeroth LLs splits into two sublevels with opposite spin states, while the LL$_{-0}$ and LL$_{+0}$ are spin-polarized and have spin-up and -down state, respectively (3, 29). The energy of the sublevel has the form (40):

$$E_{N,\xi} = E_N(k_z = 0) + 1/2\xi g_N B \tag{4}$$

where $\xi$ is equal to +1 for spin-up and –1 for spin-down and $g_N$ is the effective Landé g-factor of LL$_N$. The spin-orbit coupling (SOC) in ZrTe$_5$ mixes the spin states of the two sublevels, so two extra optical transitions between the sublevels with different spin-indices become possible. The inter- and intra-LL transition energies including the Zeeman effect can be written as (40):

$$E_{N,\xi,\xi'}^{Inter} = E_N^{Inter} + 1/2(\xi g_N - \xi' g_{-(N-1)})B \tag{5}$$

$$E_{N,\xi,\xi'}^{Intra} = E_N^{Intra} + 1/2(\xi g_N - \xi' g_{(N-1)})B \tag{6}$$

where $\xi$ and $\xi'$ correspond to the spin states of the two sublevels, respectively.

Figure 3*B* displays the false-color map of the $-\ln[T(B)/T(B_0)]$ spectra of the ZrTe$_5$ flake. Interestingly, a cusp-like feature around 18 T, which is indicated by a white arrow, can be observed in Fig. 3*C* (i.e. the magnified image of a region in Fig. 3*B*). To quantitatively investigate the physical meaning of this cusp-like feature, we plot the energies of the four split T$_1$ transitions (i.e. 1$\alpha$, 1$\beta$, 1$\gamma$ and 1$\delta$ (green dots)) around 16 T in Fig. 3*B*, which are defined by the onsets of the absorption features due to the Zeeman splitting (see Fig. 3 of the Supplemental Material of Ref. 40 and Supplementary Section 6). As displayed

by the green dashed lines in Fig. 3B, fitting the energy traces of the inter-LL transitions, $1\alpha$, $1\beta$, $1\gamma$ and $1\delta$, based on Eq. (5) with the obtained values of the Fermi velocity $v_F^{T_1}$ and the bandgap $\Delta$ yields the g-factors of the two zeroth LLs and $LL_{\pm 1}$: $g_{eff}(LL_{+0}) = g_{eff}(LL_{-0}) \approx 11.1$, $g_{eff}(LL_{-1}) \approx 31.1$ and $g_{eff}(LL_{+1}) \approx 9.7$ (or $g_{eff}(LL_{+1}) \approx 31.1$ and $g_{eff}(LL_{-1}) \approx 9.7$) (see Supplementary Section 5).

It is known that as a hallmark of TIs, the band inversion causes the exchange of the characteristics between the valence- and conduction-band extrema (2, 6, 28), so as shown in Fig. 1A and 3D, the $LL_{-0}$ and $LL_{+0}$, which come from the inverted band extrema, have reversed spin states and cross at a critical magnetic field (3, 29). According to Eq. (4) with the above values of $g_{eff}(LL_{\pm 0})$ and $\Delta$, we estimated the critical magnetic field $B_c \approx 17$ T. In Fig. 2A and D, the disappearance of the intra-LL transition $T_1^*$ around $B \approx 2.5$ T indicates that $LL_{+0}$ (or $LL_{-0}$) becomes fully depleted (or occupied) with increasing magnetic field and that at $B > 2.5$ T, the chemical potential of $ZrTe_5$ can be considered to be located at zero energy. In this case, the two zeroth LLs intersect with the chemical potential at the same magnetic field $B_c$. More importantly, this intersection means at $B > B_c \approx 17$ T, $LL_{-0}$ and $LL_{+0}$ becomes empty and occupied, respectively, which leads to the gradual replacement of the inter-LL transitions $T_1$, $1\alpha$, $1\beta$, $1\gamma$ and $1\delta$, by the intra-LL transitions $T_1^*$, $1\chi$, $1\lambda$, $1\theta$ and $1\varepsilon$, explained in Fig. 3D. Furthermore, in Fig. 3B, the energy traces of the four split transitions (grey dots) observed at $B > 17$ T are shown to follow the white theoretical curves for the intra-LL transitions $T_1^*$, which are based on Eq. (6). Therefore, the four split transitions observed at $B > 17$ T in Fig. 3B can be assigned as the intra-LL transitions $T_1^*$, $1\chi$, $1\lambda$, $1\theta$ and $1\varepsilon$. Since the energy traces of the split intra-LL transitions $T_1^*$ deviate markedly from those of the inter-LL transitions $T_1$, the reemergence of the $T_1^*$ transitions at $B > 17$ T causes the cusp-like feature, which provides experimental evidence for the bulk-band inversion in the $ZrTe_5$ flake.

In summary, using magneto-infrared spectroscopy, we have investigated the Landau level spectrum of the multilayer $ZrTe_5$ flake. The magnetic-field-dependence of the LL transition energies here, together with the photon-energy-dependence of the absorption coefficient at zero-field, quantitatively demonstrates 3D massive Dirac fermions with nearly linear dispersions in the $ZrTe_5$ flake. Due to the Zeeman splitting of the LLs, the energy splitting of the LL transitions was observed at $B \geq 6$ T. Interestingly, the intra-LL transitions $T_1^*$ reemerge at $B > 17$ T. We propose that the reemergence of the $T_1^*$ transitions results from the band-inversion-induced crossing of the two zeroth LLs, $LL_{+0}$ and $LL_{-0}$. Our results make $ZrTe_5$ flakes good contenders for 3D TIs. Moreover, due to the 3D massive Dirac-like dispersions and the high bulk-carrier mobility ($\approx 33,000$ cm$^2$/ (V·s)), the $ZrTe_5$ flake can also be viewed as a 3D analogue of gapped graphene, which enables us to deeply investigate exotic quantum phenomena.

## Materials and Methods

**Sample preparation and characterizations.** Bulk single crystals of $ZrTe_5$ were grown by Te flux method. The elemental Zr and Te with high purity were sealed in an evacuated double-walled quartz ampule. The raw materials were heated at 900 ℃ and kept for 72 hours. Then they were cooled slowly down to 445 ℃ and heated rapidly up to 505 ℃. The thermal-cooling cycling between 445 and 505 ℃ lasts for 21 days to re-melt the small size crystals. The multilayer $ZrTe_5$ flake (*ac*-plane) for magneto-transmission measurements were fabricated by mechanical exfoliation, and deposited onto double-side-polished $SiO_2$/Si substrates with 300 nm $SiO_2$. The flake thickness ~ 180 nm and the chemical composition were characterized by atomic force microscopy (AFM) and energy dispersion spectroscopy (EDS), respectively (Please see Fig. S1).

**Infrared transmission measurements.** The transmission spectra were measured at about 4.5 K in a resistive magnet in the Faraday geometry with magnetic field applied in parallel to the wave vector of incident light and the crystal *b*-axis. Non-polarized IR light (provided and analyzed by a Fourier transform spectrometer) was delivered to the sample using a copper light pipe. A composite Si bolometer was placed directly below the sample to detect the transmitted light. The diameter of IR focus on the sample is ∼ 0.5 – 1 mm. Owing to the mismatch between the size of the IR focus and the $ZrTe_5$ flake, an aluminium aperture was placed around the sample. The transmission spectra are shown at energies above 10 meV, corresponding to wavelengths shorter than 124 μm. The wavelength of infrared light here is smaller than the size of the measured sample, and thus the optical constants can be used for a macroscopic description of the data.

## References

1. Kane CL, Mele EJ. (2005) $Z_2$ topological order and the quantum spin Hall effect. *Phys Rev Lett* 95: 146802.
2. Bernevig BA, Hughes TL, Zhang SC. (2006) Quantum spin Hall effect and topological phase transition in HgTe quantum wells. *Science* 314:1757–1761.
3. König M, et al. (2007) Quantum spin Hall insulator state in HgTe quantum wells. *Science* 318:766–770.
4. Fu L, Kane CL, Mele EJ, (2007) Topological insulators in three dimensions. *Phys Rev Lett* 98:106803.
5. Moore JE, Balents L, (2007) Topological invariants of time-reversal-invariant band structures. *Phys Rev B* 75:121306.
6. Zhang H, et al. (2009) Topological insulators in $Bi_2Se_3$, $Bi_2Te_3$ and $Sb_2Te_3$ with a single Dirac cone on

the surface. *Nat Phys* 5:438–442.

7. Hsieh D, et al. (2008) A topological Dirac insulator in a quantum spin Hall phase. *Nature* 452:970–974.

8. Chen YL, et al. (2009) Experimental realization of a three-dimensional topological insulator, $Bi_2Te_3$. *Science* 325:178–181.

9. Wan XG, Turner AM, Vishwanath A, Savrasov SY, (2011) Topological semimetal and Fermi-arc surface states in the electronic structure of pyrochlore iridates. *Phys Rev B* 83**:**205101.

10. Young SM, et al. (2012) Dirac semimetal in three dimensions. *Phys Rev Lett* 108**:**140405.

11. Wang ZJ, et al. (2012) Dirac semimetal and topological phase transitions in $A_3Bi$ (A = Na, K, Rb). *Phys Rev B* 85**:**195320.

12. Wang ZJ, et al. (2013) Three-dimensional Dirac semimetal and quantum transport in $Cd_3As_2$. *Phys Rev B* 88**:**125427.

13. Liu ZK, et al. (2014) Discovery of a three-dimensional topological Dirac semimetal, $Na_3Bi$. *Science* 343**:**864–867.

14. Neupane M, et al. (2014) Observation of a three-dimensional topological Dirac semimetal phase in high-mobility $Cd_3As_2$. *Nat Commun* 5:3786.

15. Jeon S, et al. (2014) Landau quantization and quasiparticle interference in the three-dimensional Dirac semimetal $Cd_3As_2$. *Nat Mater* 13:851–856.

16. Liang T, et al. (2014) Ultrahigh mobility and giant magnetoresistance in the Dirac semimetal $Cd_3As_2$. *Nat Mater* 14:280–284.

17. He LP, et al. (2014) Quantum transport evidence for a three-dimensional Dirac semimetal phase in $Cd_3As_2$. *Phys Rev Lett* 113:246402.

18. Borisenko S, et al. (2014) Experimental realization of a three-dimensional Dirac semimetal. *Phys Rev Lett* 113:027603.

19. Liu ZK, et al. (2014) A stable three-dimensional topological Dirac semimetal $Cd_3As_2$. *Nat Mater* 13: 677–681.

20. Burkov AA, Balents L, (2011) Weyl semimetal in a topological insulator multilayer. *Phys Rev Lett* 107:127205.

21. Weng H, et al. (2015) Weyl semimetal phase in non-centrosymmetric transition metal monophosphides. *Phys Rev X* 5:011029.

22. Huang SM, et al. (2015) An inversion breaking Weyl semimetal state in the TaAs material class. *Nat Commun* 6:7373.

23. Xu SY, et al. (2015) Discovery of a Weyl Fermion semimetal and topological Fermi arcs. *Science* 347: 294.

24. Lv BQ, et al. (2015) Observation of Weyl nodes in TaAs. *Nat Phys* 11:724–727.

25. Yang LX, et al. (2015) Weyl semimetal phase in the non-centrosymmetric compound TaAs. *Nat Phys* 11:728–732.

26. Xu SY, et al. (2015) Discovery of a Weyl fermion state with Fermi arcs in niobium arsenide. *Nat Phys.* 11:748–754.

27. Shekhar C, et al. (2015) Extremely large magnetoresistance and ultrahigh mobility in the topological Weyl semimetal candidate NbP. *Nat Phys* 11:645–649.

28. Liu CX, et al. (2010) Model Hamiltonian for topological insulators. *Phys Rev B* 82:045122.

29. Orlita M, et al. (2015) Magneto-optics of massive Dirac fermions in bulk $Bi_2Se_3$. *Phys Rev Lett* 114:186401.

30. Hunt B, et al. (2013) Massive Dirac fermions and Hofstadter butterfly in a van der Waals heterostructures. *Science* 340:1427.

31. Chen Z-G, et al. (2014) Observation of an intrinsic bandgap and Landau level renormalization in graphene/boron-nitride heterostructures. *Nat Commun* 5:4461.

32. Fjellvåg H, Kjekshus A, (1986) Structural properties of $ZrTe_5$ and $HfTe_5$ as seen by powder Diffraction. *Solid State Commun* 60:91.

33. Weng H, et al. (2014) Transition-Metal Pentatelluride $ZrTe_5$ and $HfTe_5$: A paradigm for large-gap quantum spin Hall insulators. *Phys Rev X* 4:011002.

34. Niu J, et al. (2015) Electrical transport in nano-thick $ZrTe_5$ sheets: from three to two dimensions. Preprint at <http://arxiv.org/abs/1511.09315>.

35. Li X-B, et al. (2016) Experimental Observation of Topological Edge States at the Surface Step Edge of the Topological Insulator $ZrTe_5$. *Phys Rev Lett* 116:176803.

36. Wu R, et al. (2016) Experimental evidence of large-gap two-dimensional topological insulator on the surface of $ZrTe_5$. *Phys Rev X* 6:021017.

37. Zhang Y, et al. (2016) Electronic evidence of temperature-induced Lifshitz transition and topological nature in $ZrTe_5$. Preprint at <http://arxiv.org/abs/1602.03576>.

38. Li Q, et al. (2016) Chiral magnetic effect in $ZrTe_5$. *Nat Phys* 12:550.

39. Chen RY, et al. (2015) Optical spectroscopy study of the three-dimensional Dirac semimetal $ZrTe_5$. *Phys Rev B* 92:075107.

40. Chen RY, et al. (2015) Magnetoinfrared spectroscopy of Landau levels and Zeeman splitting of three-dimensional massless Dirac fermions in $ZrTe_5$. *Phys Rev Lett* 115:176404.

41. Zheng G, et al. (2016) Transport evidence for the three-dimensional Dirac semimetal phase in $ZrTe_5$. *Phys Rev B* 93:115414.

42. Orlita M, et al. (2014) Observation of three-dimensional massless Kane fermions in a zinc-blende

crystal. *Nat Phy* 10:233.

43. Li ZQ, et al. (2008) Dirac charge dynamics in graphene by infrared spectroscopy. *Nat Phys* 4:532-535.

44. Sadowski ML, et al. (2006) Landau level spectroscopy of ultrathin graphite layers. *Phys Rev Lett* 97:266405.

45. Jiang Z, et al. (2007) Infrared spectroscopy of Landau levels of graphene. *Phys Rev Lett* 98:197403.

46. Liang T, et al. (2013) Evidence for massive bulk Dirac fermions in $Pb_{1-x}Sn_xSe$ from Nernst and thermopower experiments. *Nature Commun* 4:2696.

47. Issi J-P, et al. (2014) Electron and phonon transport in graphene in and out of the bulk. *Physics of Graphene*, eds Aoki H, Dresselhaus MS (Springer), pp 65-112.

48. Kamm GN, et al. (2014) Fermi surface, effective masses, and Dingle temperatures of $ZrTe_5$ as derived from the Shubnikov–de Haas effect. *Phys Rev B* 31:7617.

49. Yuan X, et al. (2015). Observation of quasi-two-dimensional Dirac fermions in $ZrTe_5$. Preprint at <http://arxiv.org/abs/1510.00907>.

50. Ponomarenko LA, et al. (2013) Cloning of Dirac fermions in graphene superlattices. *Nature* 497:594–597.

51. Dean CR, et al. (2013) Hofstadter's butterfly and the fractal quantum Hall effect in moiré superlattices. *Nature* 497:598-602.

## Acknowledgments

The authors thank X. C. Xie, F. Wang, Z. Fang, M. Orlita, M. Potemski, H. M. Weng, L. Wang, C. Fang and X. Dai for very helpful discussions. The authors acknowledge support from CAS Pioneer Hundred Talents Program, the National Key Research and Development Program of China (Project No. 2016YFA0300600), the European Research Council (ERC ARG MOMB Grant No. 320590), the National Science Foundation of China (Grant No. 11120101003 and No. 11327806) and the 973 project of the Ministry of Science and Technology of China (Grant No. 2012CB821403). A portion of this work was performed in National High Magnetic Field Laboratory which is supported by National Science Foundation Cooperative Agreement No. DMR-1157490 and the State of Florida. Work at Brookhaven was supported by the Office of Basic Energy Sciences (BES), Division of Materials Sciences and Engineering, U.S. Department of Energy (DOE), through Contract No. DE-SC00112704.

## Author Contributions

Z.G.C. conceived this research project, carried out the optical experiments and wrote the paper. Z.G.C., N.L.W., R.Y.C. and R.Y. analyzed the data. G.D.G., R.D.Z. and J.S. grew the single crystals. Q.L., C.Z.,

Y.H. and F.Q. performed the basic characterization. All authors discussed the results and commented on the paper.

## Figure Legends

**Fig. 1.** Bulk-band and crystal structure of ZrTe$_5$. (**A**) Top row: schematic of the topological phase transition from trivial to topological insulators (TIs). A 3D Dirac semimetal (DS) can be regarded as a quantum critical point with a gapless band structure. Due to the bulk-band inversion in TIs, the conduction and valence band exchange their extrema. Bottom row: energy spectrum for the Landau-index $N = +0$ and $-0$ LLs with a Zeeman splitting. (**B**) Atomic structure of ZrTe$_5$. Each unit cell contains two ZrTe$_5$ layers. (**C**) Left: AFM image of ZrTe$_5$ flakes. Right: thicknesses along the colored lines on the left. Thicknesses from 5 to 13 unit cells (u.c.) are shown. (**D**) Absorption coefficient $A(\omega)$ of the ZrTe$_5$ flake with thickness $d \sim 180$ nm as a function of photon energy at $B = 0$ T. The linear energy-dependence of $A(\omega)$ is mainly associated with inter-band transitions of 3D Dirac electrons. A sudden drop in $A(\omega)$ at low energies implies the modification of the band structures within an energy range, namely the bandgap. The inset of (**D**) depicts the inter-band absorption in gapped ZrTe$_5$.

**Fig. 2.** Low-magnetic-field Landau level transitions in ZrTe$_5$. (**A**) Absorption features T$_n$ ($1 \leq n \leq 7$) in $T(B)/T(B_0)$ spectra. The spectra are displaced from one another by 0.1 for clarity. (**B** and **C**) Squares of T$_n$ and T$_1^*$ energies plotted as a function of magnetic field ($B \leq 4$ T). T$_n$ correspond to the interband LL transitions: LL$_{-|N|} \rightarrow$ LL$_{+|N-1|}$ (or LL$_{-|N-1|} \rightarrow$ LL$_{+|N|}$). The non-zero $E_{T_1}^2(B)$ at zero-field reveals a bandgap $\Delta$. The upper left inset of (**C**) shows the energy difference between T$_1$ and T$_1^*$, which equals to the bandgap. (**D**) Absorption features T$_1^*$ in $T(B)/T(B_0)$ spectra. T$_1^*$, which is located at energies lower than T$_1$, arises from the intraband LL transition LL$_{+0} \rightarrow$ LL$_{+1}$ (or LL$_{-1} \rightarrow$ LL$_{-0}$). (**E**) Schematic of the Landau level spectrum without a Zeeman splitting and the allowed optical transitions, shown in a $E^2$-$B$ plot.

**Fig. 3.** Crossing of the two zeroth LLs of ZrTe$_5$. (**A**) Split T$_1$ transitions in the $T(B)/T(B_0)$ spectra at $B \geq 6$ T. The energies of the split T$_1$ transitions are defined by the onsets of the dip features indicated by the blue triangles. Around 16 T, four modes are present. Four modes reemerge around 27 T. (**B**) Color scale map of the $-\ln[T(B)/T(B_0)]$ spectra as a function of magnetic field and energy. (**C**) Magnified view of a region in (**B**) to better present the cusp-like feature which is indicated by a white arrow around 18 T. The measured T$_1$ (green dots) and T$_1^*$ (grey dots) energies are plotted as a function of $B$ in (**B**). Here, the green and grey dots in (**B**) can be extracted from the lower-energy edges (i.e. the onsets) of the peak-feature traces in the color intensity plots (i.e. (**B**) and (**C**)) of the $-\ln[T(B)/T(B_0)]$ spectra. These dots in (**B**) have the intensities of the color scale, respectively: 0.025 for 1$\alpha$, 1$\beta$, 1$\gamma$ and 1$\theta$, –0.025 for 1$\delta$, 1$\chi$ and 1$\lambda$ and –0.075 for 1$\varepsilon$. The theoretical T$_1$ and T$_1^*$ energies, based on Eq. (5) and (6) with the g-factors: $g_{\text{eff}}(\text{LL}_{-1}) = 31.1$, $g_{\text{eff}}(\text{LL}_{+1}) = 9.7$, and $g_{\text{eff}}(\text{LL}_{+0}) = g_{\text{eff}}(\text{LL}_{-0}) = 11.1$, the bandgap $\Delta = 10$ meV, the Fermi velocities $v_F^{T_1} = 4.76 \times 10^5$ m/s and $v_F^{T_1^*} = 4.63 \times 10^5$ m/s, are shown by the green and grey dashed curves, respectively. (**D**) Schematic of the split T$_1$ (green arrows) and T$_1^*$ (grey arrows) transitions. The LL spectrum is produced with the above values of the g-factors, the bandgap and the Fermi velocity $v_F^{T_1}$. The two zeroth LLs cross at a critical magnetic field $B_c \approx 17$ T. The chemical potential is roughly at zero energy when the magnetic field is high enough. At $B > B_c$, the interband LL transitions (1$\alpha$, 1$\beta$, 1$\gamma$ and $1\delta$) are gradually replaced by the intraband LL transitions ($1\chi$, $1\theta$, $1\lambda$ and $1\varepsilon$), which causes the cusp-like feature shown in (**C**).

# Figures

**Fig. 1**

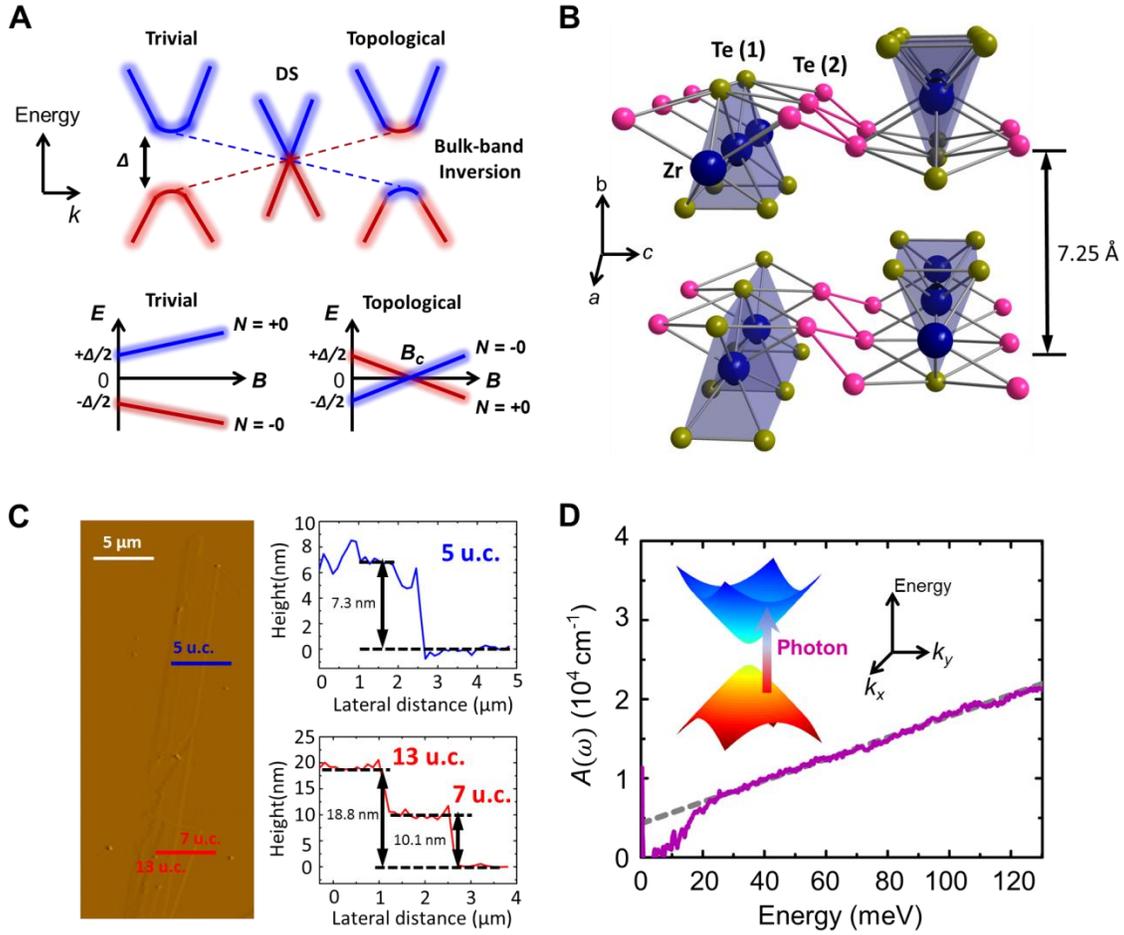

**Fig. 2**

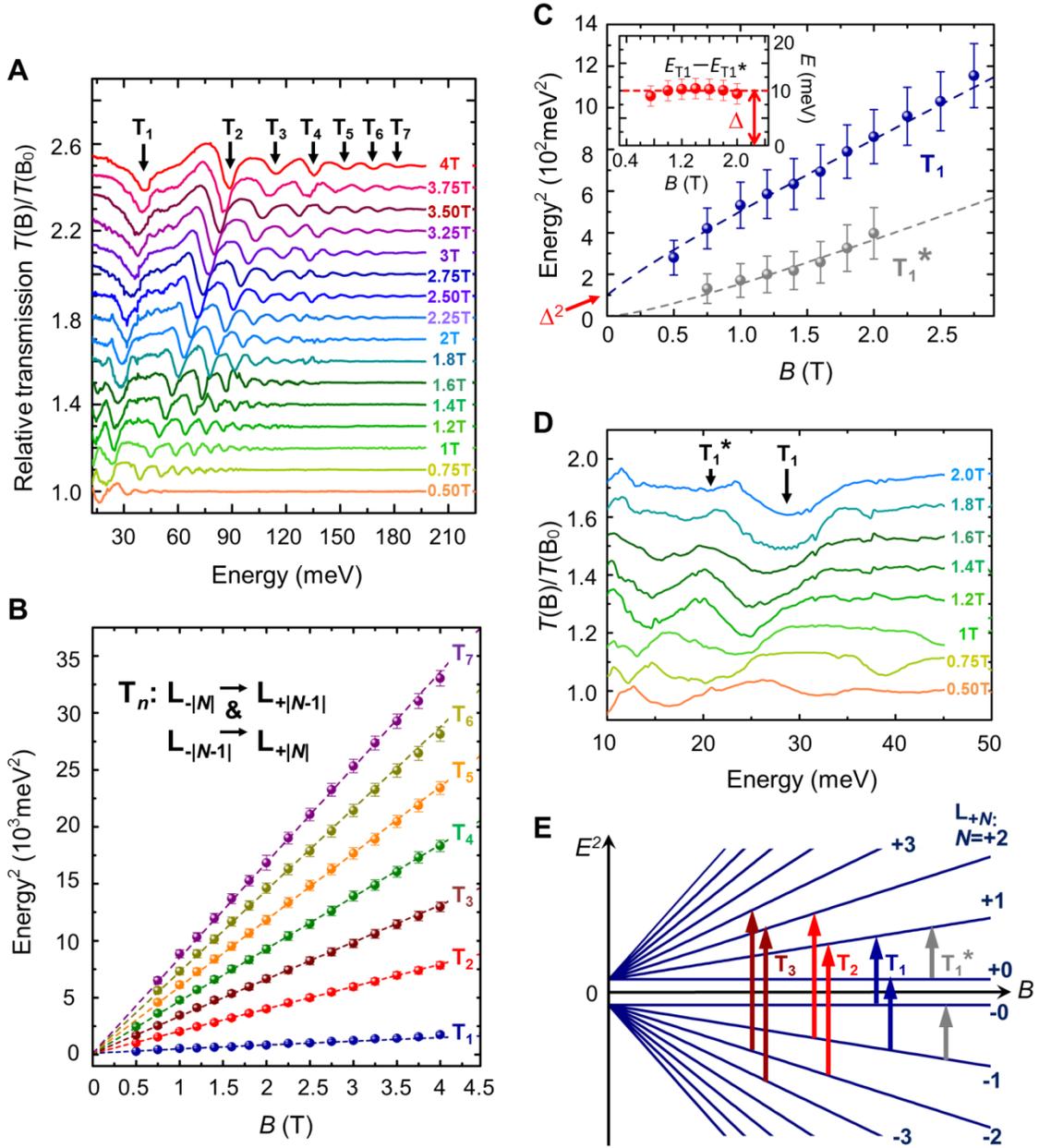

**Fig. 3**

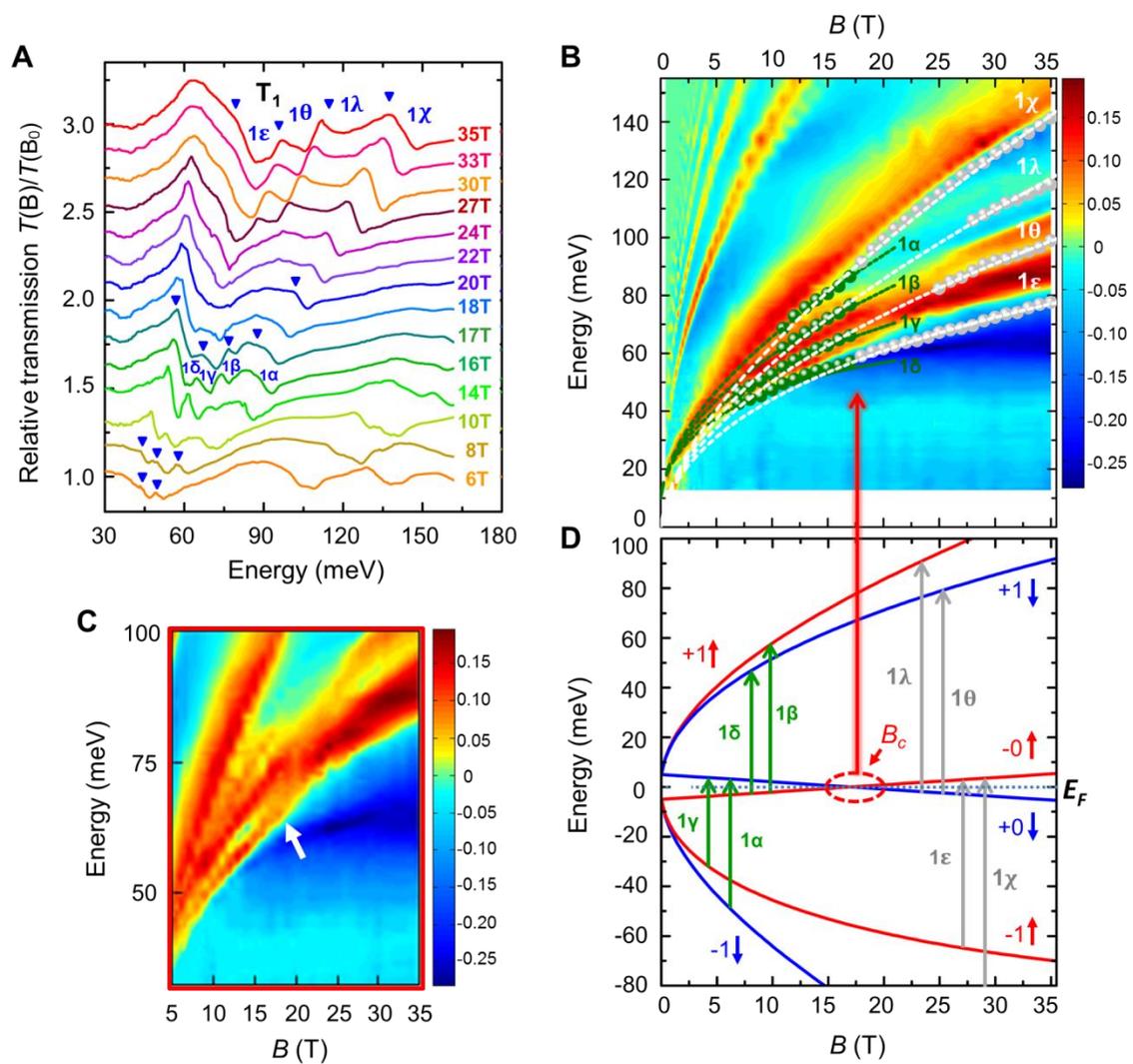